\documentclass[useAMS,usenatbib]{mn2e}
\usepackage[latin1]{inputenc}
\usepackage{graphicx}
\usepackage{amssymb}
\usepackage{mathrsfs}
\usepackage[usenames,dvipsnames]{xcolor}
\usepackage{natbib}
\bibpunct{(}{)}{;}{a}{}{,}

\title[A nonextensive approach to the stellar rotational evolution I]{A
nonextensive approach to the stellar rotational evolution I. F and G type stars}
\author[D. B. de Freitas and J. R. De Medeiros]{D. B. de
Freitas$^{1}$\thanks{E-mail:
danielbrito@dfte.ufrn.br} and J. R. De Medeiros$^{1}$\\
$^{1}$Departamento de F\'{\i}sica,
    Universidade Federal do Rio
    Grande do Norte, 59072-970
    Natal,  RN, Brazil}
\begin{document}

\date{Accepted --. Received --; in original
form --}

\pagerange{\pageref{firstpage}--\pageref{lastpage}} \pubyear{2002}

\maketitle

\label{firstpage}

\begin{abstract}
The pioneering study by Skumanich (1972) showed that the rotational 
velocity of G-type
Main-Sequence (MS) stars decreases with stellar age according to $\left\langle v \sin i\right\rangle$
$\propto$ $t^{-1/2}$.
This relationship is consistent with simple theories of angular momentum 
loss from rotating stars,
where an ionized wind is coupled to the star by a magnetic field. The 
present study introduces a new
approach to the study of stellar rotational braking in unsaturated F and G type 
stars limited in age and mass, connecting angular momentum loss
by magnetic stellar wind with Tsallis nonextensive statistical mechanics.
As a result, we show that the rotation-age relationship can be well 
reproduced using a nonextensive approach
from Tsallis nonextensive models. Here, the index $q$, which is 
related to the degree of nonextensivity,
can be associated to the dynamo process and to magnetic field geometry, 
offering relevant information on
the level of stellar magnetic braking for F- and G-type Main-Sequence  
stars.
\end{abstract}


\section{Introduction}
The relationship between stellar rotation and age is an exciting topic in
astrophysics posing, a number of unanswered questions, including: 1) How is
the stellar rotation-age relationship dependent on mass? 2) How does
rotation influence the stellar dynamo, as well as chromospheric and
coronal heating mechanisms? 3) How do initial star formation systems
affect rotational evolution? 4) How do stellar angular momentum and
stellar activity evolve over time?

\citet{Skumanich}'s pioneering study showed that $v$ sin$i$ of G-type
Main-Sequence (MS) stars for Hyades and Pleiades measured by \citet{Kraft}
were consistent with $\left\langle v \sin i\right\rangle$ $\propto$ $t^{-1/2}$, where $t$ is the stellar age,  $v$ is the equatorial rotational velocity of the star and $i$ the inclination angle of the rotational axis to the line of sight.
This relationship is in
line with simple theories of angular momentum loss from rotating stars,
where an ionized wind is coupled to the star by a magnetic field
\citep{Schatzman}. Indeed, angular momentum loss due to stellar winds is
generally believed to be responsible for the Skumanich relationship, but
the exact dependence of rotation on age has not been fully established
\citep{kawaler1988,kris1997}. \citet{Barry} and \citet{Soderblom1991}
reported similar qualitative results for solar-type stars, but with
power-law exponents ranging from $-1/2$ (corresponding to the Skumanich
relation) to $-4/3$. \citet{pacepas} claimed that these
power-laws do not fit the age-activity-rotation of G dwarf stars in open
clusters. According to these authors, a $t^{-5/2}$ law is more consistent
with the observations. Despite the differences in
rotation-activity-age relationships obtained to date, most of the data strongly
suggest that this relationship is deterministic and not merely a statistical
artifact. Indeed, rotation data for stars at different evolutionary stages and 
environments, in particular for stars in open clusters, have confirmed the 
deterministic nature (e.g.: \citep{barnes2010,reiners2012}\citet{barnes2003}, \citet{mamajek2008}, \citet{james2010} and \citet{meibom2011,meibom2011b}.

In this study we present a new approach to the study of stellar
rotational braking in F and G type stars, connecting angular momentum loss
by magnetic stellar wind \citep{kawaler1988,chaboyer1995} with Tsallis'
nonextensive statistical mechanics \citep{tsallis1988}. In Section 2, we revisit parametric models for angular
momentum loss by magnetic stellar wind, with an emphasis on a modified
Kawaler model. Section 3 describes our nonextensive approach, in Section 4 we describe the observational data, and Section
5 compares our model with $v$ sin$i$ measurements for a large sample of F-
and G-type single main--sequence stars. Finally, conclusions are presented in Section 5.

\section{Revisiting angular momentum loss by magnetic stellar wind}

Following the theory of angular momentum loss elaborated by
\citet{mestel1968,mestel1984} and \citet{mestel1987}, \citet{kawaler1988} developed a general
parametrization showing that the rate of angular momentum loss is
proportional to $\Omega^{1+4aN/3}$; where $\Omega$ is angular velocity,
$a$ denotes the dynamo relationship and $N$ is a measurement of magnetic
field geometry. The problem with angular momentum loss due to magnetic
stellar wind is its dependence on these two parameters, $a$ and $N$.
\citet{kawaler1988} simplified the dynamo relationship by assuming that
the total magnetic field is proportional to the rotation rate to some
power, denoted by $B_{0}\propto \Omega^{a}$, where $a$ varies between 1
and 2 for the unsaturated domain and 0 for the saturated domain. As
demonstrated by \citet{chaboyer1995}, the $N$ parameter exhibits a wide
range, varying from 0 to 2, where 0 represents dipole field geometry and 2
a radial field. Moreover, these authors have shown that angular momentum
loss from high rotation stars cannot be described by Kawaler's
relationship. The same authors and subsequent studies \citep{reiners2012}
have also demonstrated that magnetic fields at high rotation rates must
saturate, following the form $\mathrm dJ/\mathrm dt \propto \Omega$.  In addition, studies by Matt and Pudritz \citep{matt2005,matt2008,matt2008b} have explored the role of stellar winds to the spin-up and spin-down torques expected to arise from the magnetic interaction between a slowly rotating pre-main-sequence star and its accretion disk, demonstrating that if stellar mass outflow rates are substantially greater than the solar rate, the stellar winds can carry off orders of magnitude with more angular momentum than can be transferred to the disk.

In terms of parametrization, there are two parameterized forms known as modified Kawaler's
relationships:

\begin{eqnarray}
\label{loss2}
\frac{\mathrm dJ}{\mathrm
dt}=-K_{w}\Omega\omega^{4aN/3}_{sat}\left(\frac{R}{R_{\odot}}\right)^{2-N}\cdot\nonumber\\
\cdot\left(\frac{M}{M_{\odot}}\right)^{-N/3}\left(\frac{\dot{M}}{10^{-14}}\right)^{1-2N/3}
\end{eqnarray}

for $\Omega\geq\omega_{sat}$ and

\begin{eqnarray}
\label{loss1}
\frac{\mathrm dJ}{\mathrm
dt}=-K_{w}\Omega^{1+4aN/3}\left(\frac{R}{R_{\odot}}\right)^{2-N}\cdot\nonumber\\
\cdot\left(\frac{M}{M_{\odot}}\right)^{-N/3}\left(\frac{\dot{M}}{10^{-14}}\right)^{1-2N/3}
\end{eqnarray}

for $\Omega<\omega_{sat}$.

Here, $\omega_{sat}$ is the threshold angular velocity beyond which
saturation occurs, $K_{w}$ is a calibration constant and $R$, $M$ and
$\dot{M}$ represent the stellar radius, mass and mass-loss rate in units
of $10^{-14}M_{\odot}$yr$^{-1}$, respectively. In this regard,
$\omega_{sat}$ completes the triplet-order parameter.  In fact, this saturation threshold is an important ingredient for theoretical models of stellar angular momentum evolution. For instance, \citet{kris1997} and \citet{sills2000} suggest that the saturation threshold depends on stellar mass and is inversely proportional to the global convective overturn timescale, which is related to the Rossby number, an indicator of the strength of dynamo magnetic activity. By contrast, \citet{reiners2012} argue that $\omega_{sat}$ does not depend on mass. According to \citet{pizzolato2003}, the saturation phenomenon is also based on magnetic proxies and its level can be inferred, irrespective of angular momentum evolution arguments, which, do not require a Rossby number for either regime. These authors also argue that the X-ray emission level from saturated stars depends only on bolometric luminosity, whereas for non-saturated stars the emission level depends only on the rotational period. This implies a direct mass-dependent $\omega_{sat}$ \citep{kris1997}. Since this timescale does not change much in our model, we neglect its effect for a given mass \citep{barnes1996,kris1997,pizzolato2003}.

In the present study we assume that since the moment of inertia $I$ and
stellar radius $R$ changes slowly during MS evolution, the angular
momentum loss law can be simplified by considering the limit of $\mathrm
dJ/\mathrm dt$ for constants $I$ and $R$, that is, this loss law is fully specified by the rotational velocity, while the star is braked by the stellar wind \citep{bouvier1997}. We also consider that in the
absence of angular momentum loss, equatorial rotational velocity $v$ can
be determined by the simple condition of angular momentum conservation,
denoted by
\begin{equation}
\label{loss3}
J=\left(\frac{I}{R}\right)_{const.}v.
\end{equation}

It is important to underline that  in our model we consider a solid-body rotation, although this assumption can directly affect spin down effects for F-type stars. For a more complete model, it is necessary to take in consideration other physical aspects such as core-envelope decoupling \citep{cameron1993,allain1998}, as well as the evolutionary behavior of the moment of inertia\citep{wolff1997,zorec2012}. 

Two pairs of equations were combined to obtain the time dependence of $v$.
First, from eqs. (\ref{loss2}) and (\ref{loss3}) we have
\begin{equation}
\label{loss6}
v(t)=v_{0}\exp \left[-\frac{(t-t_{0})}{\tau_{1}}\right], \quad (t_{0}\leq t<
t_{sat})
\end{equation}
with
\begin{equation}
\label{loss6x}
\tau_{1}\equiv\left[f(\Lambda)\omega^{4aN/3}_{sat}\right]^{-1}
\end{equation}
for the saturated domain, where $f(\Lambda)$ is denoted by
\begin{eqnarray}
\label{loss7a}
f(\Lambda)=\frac{K_{w}}{I}\left(\frac{R}{R_{\odot}}\right)^{2-N}
\left(\frac{M}{M_{\odot}}\right)^{-N/3}\left(\frac{\dot{M}}{10^{-14}}\right)^{1-2N/3}
\end{eqnarray}
where $I$ denotes the moment of inertia of the star and $\Lambda$ denotes a parameter set given by $\Lambda=\left\{R,M,\dot{M},N\right\}$.

On the other hand, by combining eqs. (\ref{loss1}) and (\ref{loss3}) we
obtain the equation
\begin{eqnarray}
\label{loss4}
v(t)=v_{sat}\left[1+\frac{(t-t_{sat})}{\tau}\right]^{-3/4aN},
\quad (t\geq t_{sat})
\end{eqnarray}
with
\begin{equation}
\label{loss4xy}
\tau\equiv\left[\frac{4aN}{3}f(\Lambda)\omega^{4aN/3}_{sat}\right]^{-1}.
\end{equation}

\begin{table}
\caption{Best parameter values of our unsaturated model using
eq. (\ref{loss8a}). The values of $N$ using eq. (\ref{loss10}) also are shown.}
\label{tab1}
\renewcommand{\arraystretch}{1.5}
\begin{tabular}{cccc}
\hline \hline
Stars & $\left\langle M/M_{\odot}\right\rangle$ & $q$ & $N_{a=2}-N_{a=1}$ \\
\hline \hline
F0-F5 & 1.36 & 1.80$\pm$0.1 & $0.30 - 0.60$\\
F6-F9 & 1.22 & 1.96$\pm$0.2 & $0.36 - 0.72$\\
G0-G5 & 1.11 & 2.18$\pm$0.2 & $0.44 - 0.89$\\
G6-G9 & 0.98 & 4.38$\pm$0.6 & $1.27 - 2.53$\\
All F &     &   1.91$\pm$0.1 & $0.34 - 0.68$\\
All G &     &  2.18$\pm$0.4  & $0.44 - 0.89$\\

\hline
\end{tabular}
\end{table}

In this respect, the  $\tau_{1}$ and $\tau$ are the MS spin-down
timescales in the saturated and unsaturated regimes, respectively. The
other parameters, $t_{0}$, $v_{0}$ and $t_{sat}$, are the age at which the
star arrives on the MS, the rotational velocity at that time and
subsequent age at which the star slows down to values below the critical
velocity $v_{sat}$, respectively \citep{reiners2012}. Time $t_{sat}$ can
be determined by setting $v(t)=v_{sat}$ in eq. (\ref{loss6}). Thus, we
have
\begin{equation}
\label{tcrit}
t_{sat}=t_{0}+\tau_{1}\ln\left(\frac{v_{0}}{v_{sat}}\right).
\end{equation}

Eq. (\ref{loss6}) reveals that in the saturated regime the star slows down
exponentially from time $t_{0}$ up to velocity $v_{sat}$ in time
$t_{sat}$. Eq. (\ref{loss4}) shows that from this time onward, the
spin-down rate decreases to a power law given by $-3/4aN$. Thus, the star
remains within a factor of a few $v_{sat}$ for the remainder of its MS
lifetime.

\begin{figure*}
\begin{center}
\includegraphics[width=1.0\textwidth]{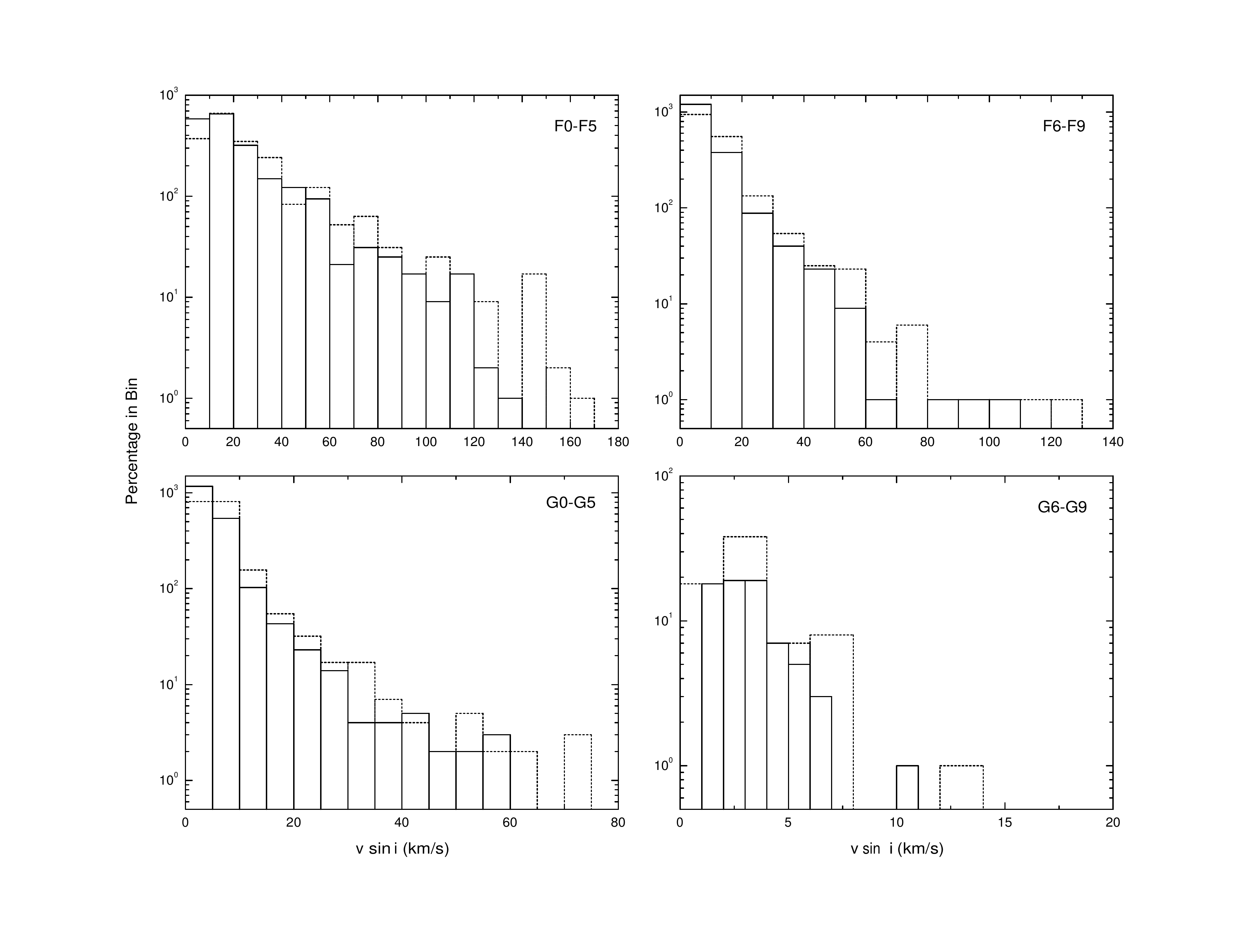}
\end{center}
\caption{
Semi-log plot of the rotational velocity distribuition for the sample of F
and G--type stars. Solid lines are refered to $v \sin i$ data, while short dashed lines are the data multiplied by the statistical correction factor $4/\pi$.}
\label{fig0}
\end{figure*}

\begin{figure*}
\begin{center}
\includegraphics[width=1.0\textwidth]{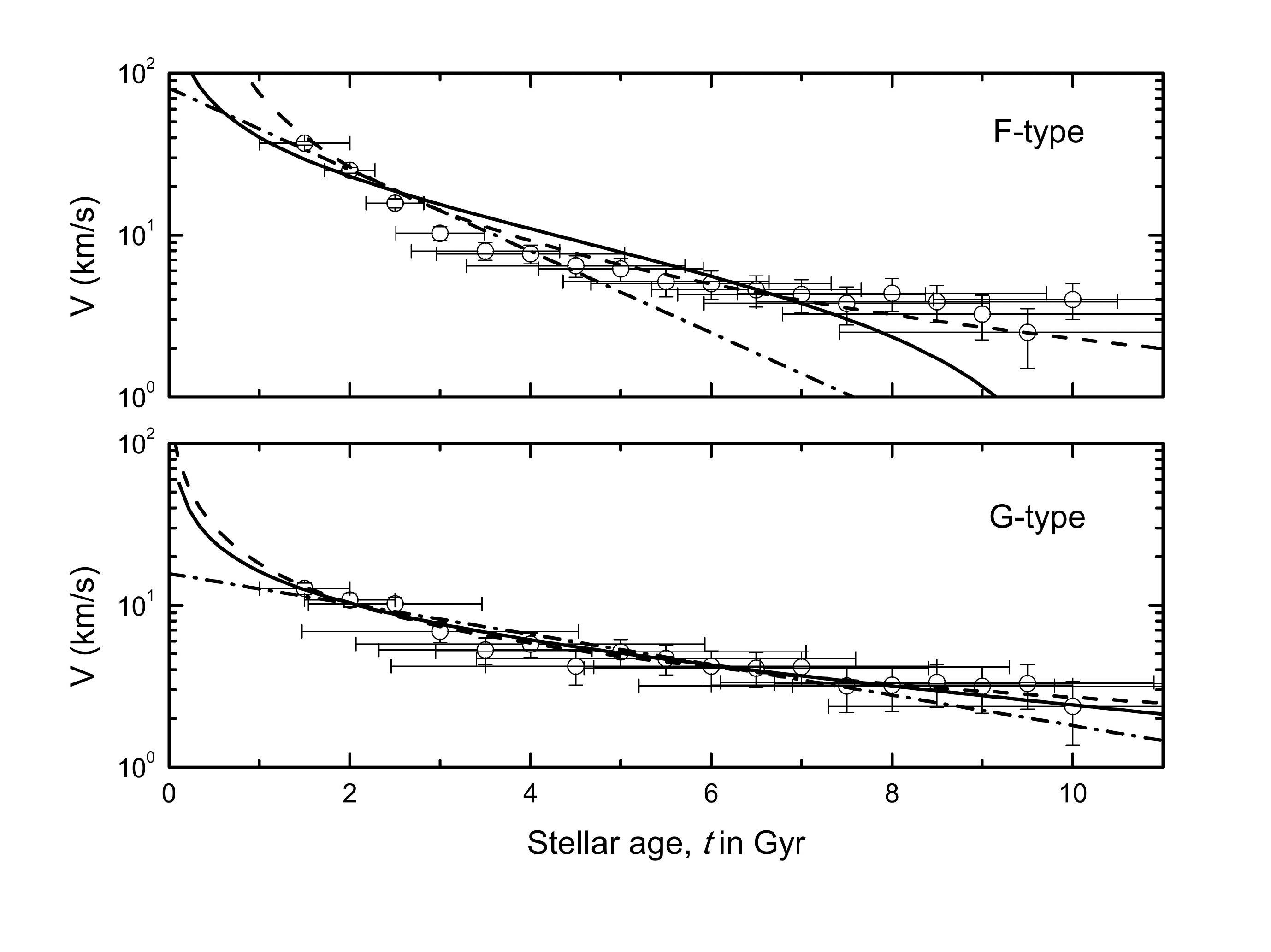}
\end{center}
\caption{
Semi-log plot of the stellar rotation-age relationship for the sample of F
and G--type stars used in the present study. Upper and lower panels
display mean projected rotational velocity, $\left\langle v \sin i\right\rangle$, by age intervals
of 0.5 Gyr for F and G-type stars, respectively, represented by the solid
circles. In both panels, continuous lines represent the fit using
Skumanich's law, and dash-dotted and dashed lines correspond to the fits
using eqs. (\ref{rela4})
and (\ref{loss8a}), respectively. Error bars for $v \sin i$ and age are also indicated.}
\label{fig1}
\end{figure*}

\begin{figure*}
\begin{center}
\includegraphics[width=1.0\textwidth]{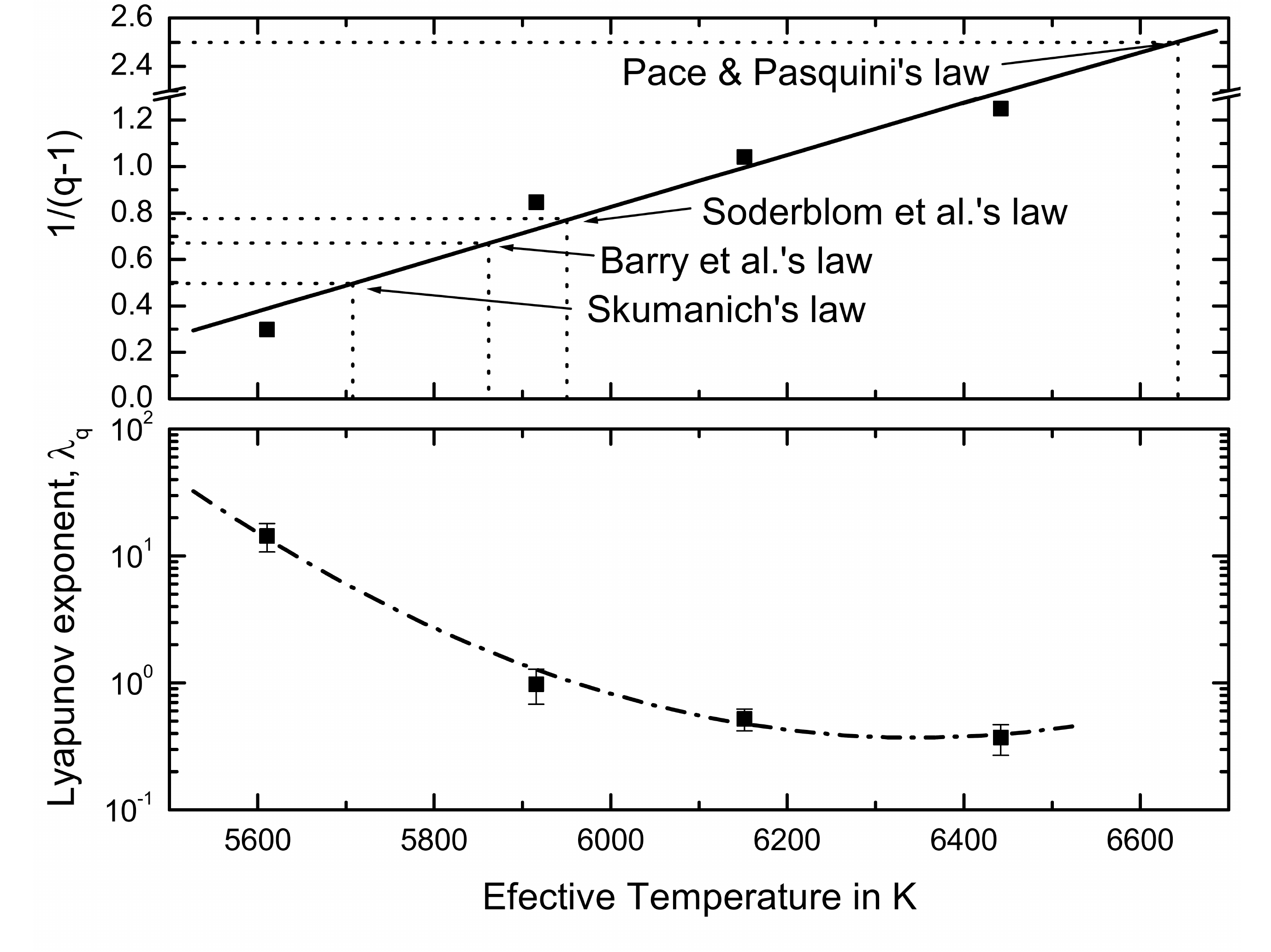}
\end{center}
\caption{The Effective Temperature dependences of
$q$ (upper panel), $\lambda_{q}$ (lower panel) used to fit the our sample segregated by different interval of spectral type. Upper figure indicates a linear dependende between the effective temperature and parameter $1/(q-1)$. For instance, Schumanich law ($1/(q-1)=0.5$) corresponds to stars with solar effective temperature. While that Pace \& Pasquini law corresponds to hottest stars.}
\label{fig2}
\end{figure*}

\section{Nonextensive approach to the stellar rotation-age relation}
This section details a nonextensive approach for the modified Kawaler
model divided into two classes: saturated and $\beta$-saturated models.
Both have the same unsaturated model, but with different saturated
timescale $t_{sat}$.

In general, for the unsaturated regime, it can be considered that the
rotation-age connection follows a Zipf-like power law \citep{zipf1949,zipf1965},
described as
\begin{equation}
\label{loss9}
\frac{v(t)}{v_{0}}=\left(1+\frac{t}{\tau}\right)^{\alpha},
\end{equation}
where $\tau$ is the timescale that controls the crossover between the
initial plateau and the power-law domain, characterized by the exponent
$\alpha$, and $v_{0}$ is a normalization constant. Equation (\ref{loss9})
was deduced within Tsallis' nonextensive theory \citep{tsallis1999,lyra},
exhibiting fractal properties and long-range memory. This theory
demonstrates that the usual exponential for saturated fields can be
recovered if $\alpha\rightarrow\infty$.

\subsection{Saturated and unsaturated regimes}
Applying Tsallis statistics \citep{tsallis2004}, we can assume that eq.
(\ref{loss6}) follows a simple linear differential equation in the form
\begin{equation}
\label{loss78}
\frac{\mathrm d}{\mathrm dt}\left(\frac{v}{v_{0}}\right)=-\lambda_{1}
\left(\frac{v}{v_{0}}\right),
\end{equation}
from which
\begin{equation}
\label{rela4}
v(t)=v_{0}\exp\left[-\lambda_{1}(t-t_{0})\right],
\end{equation}
where $\lambda_{1}$ denotes the classical Lyapunov coefficient.

According to \citet{lyra}, in contrast to the exponential behavior
presented in eq. (\ref{rela4}), eq. (\ref{loss78}) must be replaced with a
slower power law at critical points where long correlations develop. In
this context, a similar expression can be created to characterize the
possible behavior of stellar rotation distribution in an unsaturated
regime, denoted by
\begin{equation}
\label{loss8}
\frac{\mathrm d}{\mathrm
dt}\left(\frac{v}{v_{sat}}\right)=-\lambda_{q}\left(\frac{v}{v_{sat}}\right)^{q}
\quad\ (\lambda_{q}\geq 0; q\geq 1).
\end{equation}

Integrating the equation above, we have
\begin{equation}
\label{loss8a}
v(t)=v_{sat}\left[1+(q-1)\lambda_{q}(t-t_{sat})\right]^{\frac{1}{1-q}},
\end{equation}
where $\exp_{q}(x)\equiv \left[1+(1-q)x\right]^{1/(1-q)}$ denoted by $q$-exponential\footnote{See \citet{borges2004} for a review of $q$-calculus.} and $\lambda_{q}$ the nonextensive Lyapunov coefficient.

Thus, in the nonextensive scenario, eq. (\ref{loss4}) is well described as
a non-linear differential equation in form (\ref{loss8}) with solution
(\ref{loss8a}). When $q=1$, eq. (\ref{rela4}) is recovered, indicating
that a system governed by a saturated regime is in thermodynamic
equilibrium. By contrast, when $q$ differs from unity, a system controlled
by unsaturation is out of equilibrium.

In agreement with our model, for the unsaturated regime, exponent $q$ can
be described by a pair ($a,N$) given by
\begin{equation}
\label{loss10}
q\equiv1+\frac{4aN}{3},
\end{equation}
whereas for saturated regime $q=1$ due to the usual exponential function. 
From above equations, we can estimate a relationship between the characteristic time $\tau_{1}$ and $\tau_{q}$ given by
\begin{equation}
\label{relatau}
\frac{\tau_{1}}{\tau_{q}}=(q-1)\frac{\lambda_{q}}{\lambda_{1}},
\end{equation}
in this context, $\lambda$s denote the braking strength.

The $q$-parameter is related to the degree of nonextensivity that emerges
within thermodynamic formalism proposed in \citet{tsallis1988}. As
revealed in eq. (\ref{loss10}), $q$ is a function of magnetic field
topology $N$ and dynamo law $a$, which depend on stellar evolution.
According to \citet{kawaler1988}, small $N$ values result in a weak wind
that acts on the MS phase of evolution, while winds with large $N$ values
remove significant amounts of angular momentum early in the
Pre-Main-Sequence (PMS). In this phase, for a given mass, maximum
rotational velocity $v$ decreases as $N$ increases. This result is
significant because, within a thermostatistical framework in which eq.
(\ref{loss8}) naturally emerges, for a given value ($a,N$) we obtain the
scale laws found in the literature, such as those proposed by
\citet{Skumanich} and \citet{Soderblom1991}.

\subsection{$\beta$-Saturated regime}
As reported by \citet{ivanova2003}, high rotation rates can differ
significantly from the exponential decay, without which the angular
momentum loss rate necessarily saturates the magnetic field
\citep{chaboyer1995}. As such, we used a more general relationship than
that described by eq. (\ref{loss2}). Thus, this equation can be rewritten
considering that the $\mathrm dJ/\mathrm dt$ rate depends on the $\Omega$,
as follows:
\begin{eqnarray}
\label{loss2a}
\frac{\mathrm dJ}{\mathrm
dt}=-K_{w}\Omega^{\beta}\omega^{1+4aN/3-\beta}_{sat}\left(\frac{R}{R_{\odot}}\right)^{2-N}\cdot\nonumber\\
\cdot\left(\frac{M}{M_{\odot}}\right)^{-N/3}\left(\frac{\dot{M}}{10^{-14}}\right)^{1-2N/3}, \quad (t_{0}\leq t<
t_{sat}),
\end{eqnarray}
for $\beta=1$, we have the saturated domain, while for $\beta>1$ can be
describe as a \textit{quasisaturated} regime, defined here as
$\beta$-saturated. Deriving eq. (\ref{loss3}) and comparing with eq. (\ref{loss2a}), we obtain that
\begin{equation}
\label{newloss8a}
v(t)=v_{0}\left[1+\frac{(t-t_{0})}{\tau_{\beta}}\right]^{\frac{1}{1-\beta}}
\end{equation}
with
\begin{equation}
\label{newloss8ax}
\tau_{\beta}\equiv\left[(\beta-1)f(\Lambda)\left(\frac{v_{0}}{v_{sat}}\right)^{\beta-1}\omega^{4aN/3}_{sat}\right]^{-1}.
\end{equation}

Following the same procedure used to obtain $v(t)$ in the previous cases,
we have
\begin{equation}
\label{loss4x}
v(t)=v_{0}\left[1+(\beta-1)\lambda_{\beta}(t-t_{0})\right]^{\frac{1}{1-\beta}},
\end{equation}

where, $\beta$ is defined in interval $1\leq\beta<1+4aN/3$.

In this regime, a generalized expression to (\ref{tcrit}) can be obtained using
the definition of so-called $q$-logarithm: $\ln_{q}(x)=\frac{x^{1-q}-1}{1-q}$. Thus,
\begin{equation}
\label{tcrit3}
t_{sat}=t_{0}+(\beta-1)\tau_{\beta}\ln_{\beta}\left(\frac{v_{0}}{v_{sat}}\right),
\end{equation}
where for $\beta=1$ the standard logarithm is recovered in (\ref{tcrit}).

In this general case, we can extrated the rate between the time $\tau_{q}$ and $\tau_{\beta}$, where we obtain
\begin{equation}
\label{rate1}
\frac{\tau_{\beta}}{\tau_{q}}=\left(\frac{q-1}{\beta-1}\right)\frac{\lambda_{q}}{\lambda_{\beta}}
\end{equation}
where $\tau_{1}$ and $\tau_{\beta}$ are related by
\begin{equation}
\label{rate1x}
\frac{\tau_{\beta}}{\tau_{1}}=\left(\frac{1}{\beta-1}\right)\left(\frac{v_{0}}{v_{sat}}\right)^{1-\beta}
\end{equation}
if $\beta=1$, we recovered the rate between $\tau_{1}$ and $\tau_{q}$ in case of the unsaturated-saturated model.

\section{The stellar sample and observational data}
Our study is based on the kinematically unbiased and
magnitude--limited sample
of nearby F and G dwarf stars from the Geneva--Copenhagen
Survey (CGS) of the
Solar neighbourhood, volume complete to $\sim$ 40 pc,
carried out by \citet{nordstrom04} and rediscussed by \citet{holmberg2007}, which contain age,
metallicity, mass, projected rotational velocity {\it v}
sin{\it i} and kinematic
properties for about 14000 F and G dwarfs. Readers are referred to these authors for observational procedure 
and data reduction, but let us briefly discuss the quality of  projected rotational velocity
{\it v} sin{\it i}, age and mass measurements, given that these parameters are the most sensitive 
in the present study. 

For the vast majority of stars (12941 stars) the
{\it v} sin{\it i} measurements were computed from
observations carried out with the CORAVEL spectrometers\citep{baranne1979,mayor1985}, applying
\citep{benz1980,benz1984}'s calibrations. Such a procedure gives {\it v}
sin{\it i} values with a precision of 1 km s$^{-1}$, at least for
rotations lower than about 30 km s$^{-1}$ \citep{medeiros1999}.

A total of 833 stars rotation measurements were determined from
observations carried out with the digital spectrometer \citep{latham1985} at the
Harvard--Smithsonian Center for
Astrophysics, CfA, in particular for stars rotating
too rapidly for CORAVEL. For these stars,
the {\it v} sin{\it i} values were computed on the basis
of the best--fitting template spectrum \citep{nordstrom1994,nordstrom1997}. As shown by these authors,
for slowly--rotating stars, typically
for {\it v} sin{\it i} values lower than 20 km s$^{-1}$,
there is a very good agreement between data based
on CfA and CORAVEL observations. As the vast majority of the present stellar working sample have
rotation below 20 km s$^{-1}$, we can consider that data
on rotation used in the present analysis have the same quality, with a
precision of about 1 km s$^{-1}$.

We assumed that the mean value for a random distribution
of inclinations $\sin i$ is given by $4/\pi$ \citep{chandra1950}. As mentioned by \citet{reiners2012}, this correction is only necessary in a statistically averaged sense and not for comparison of single observed populations. Such a fact is valid for the present stellar sample, which is unbiased and magnitude--complete. In this sense  statistical variations in the $\sin i$ distribution should not significantly change the presence or location of the very steep transition from a large population of undetected $v \sin i$ to a similarly large population of high $v \sin i$ observed \citep{reiners2012}. Fig. \ref{fig0} shows the projected velocity distribution per bin uncorrected and corrected by factor $4/\pi$, in which no significant $\sin i$ effects can be observed.

Good age measurements are, of course, crucial when
investigating the rotation--age relationship. Most of the
previous studies
on this subject have used ages estimated from
chromospheric activity. Nevertheless, such a
procedure is not valid for stars at the age of the Sun,
because chromospheric emission essentially vanishes at
this stage.
The theoretical isochrone age is then the best choice,
although important discrepancies subsist, in particular
for old low--mass
stars. The ages given in the GCS
were then obtained on the basis of the stellar isochrone
procedure
\citet{nordstrom04}, using the Bayesian computational
technique of \citet{jorgensen2005}. A
revision of the data was carried out by
\citet{holmberg2007},
taking into account new temperature and metallicity
calibrations. According \citet{nordstrom04} and
\citet{holmberg2007} for 81$\%$ of the presumably single
stars in the original sample of the
CGS, age estimations based, on their
procedure, have errors below 50$\%$. 

\citet{nordstrom04} have also estimated stellar masses
using theoretical isochrone analysis, with individual
errors averaging
about 0.05 M$_{\sun}$, and metallicities from Strömgren
$ubvyB$ photometry. As shown by these authors, the
distribution of their estimated
metallicities obeys a Gaussian distribution with a mean of
-0.14 and dispersion 0.19 dex. The distribution of the metallicities from \citet{holmberg2007} for the 5835 single stars used in the present
analysis shows a mean value $\left\langle[Fe/H]\right\rangle=-$0.18 and
dispersion
of 0.21 dex. Clearly, one observes that our working sample
is mostly composed of solar--type stars, in spite of
the presence of a low percentage of stars with slightly
sub--solar metallicities.  Recently, \citet{casagrande2011} re-analyzed most of the parameters for the CGS via the infrared flux method (IRFM). According to these authors, this method, in comparison to previous CGS calibrations, reveals differences in measurement of effective temperature leading to a much better agreement between stars and isochrones in the HR diagram.

In order to test our nonextensive approach for stellar rotational magnetic
breaking, we used  {\it v}
sin{\it i} and ages from the referred Geneva-Copenhagen survey for F- and G-type single stars
\citep{nordstrom04,holmberg2007}, with  ages limited to 10 Gyr and masses of $0.90M_{\odot}\leq
M\leq 2.0 M_{\odot}$.  This upper limit for ages was defined by the star's lifetime on the MS and the age of the Galactic disk \citep{epstein2012}, while the upper limit for masses (or spectral type) is based on the different internal structure of the stars in our sample. These age and mass limits were chosen to avoid the
presence of stars exhibiting the most significant uncertainties for these
parameters, namely the oldest and lowest mass stars, as well as the
highest mass stars. Based on these criteria, we obtained a final working
sample of 5780 single stars consisting of 3790 F- (2050 with spectral type between F0 and F5 and 1750 between F6 and F9) and 1990 G- (1919 with spectral type between G0 and G5 and 73 between G6 and G9) type stars.

\section{Testing the model with unsaturated stars}
 First of all, stars in the present working sample can be considered to be in the unsaturated regime , indicating that the computation of the values of the parameter $\beta$ is not necessary. Indeed, all the stars are aged more than 1 Gyr, as described in Section 4. 

Different studies in the literature (\textit{e.g}: \citep{kris1997,bouvier1997,epstein2012}) use the rotation distribution of young open clusters to estimate initial conditions. Here, we follow the same approach, by considering the rotation distribution of T Tauri stars. In agreement with initial conditions, we compute the best-fit $[q,\lambda_{q}]$ doublet.
Fig. \ref{fig1} and \ref{fig2} shows the results of the present analysis, where the mean of $v\sin i$
by age intervals of 0.5 Gyr are shown for the aforementioned
sample of F and G--type stars. In Fig. \ref{fig1}, the upper and lower panels
also display the best fits applying the Skumanich law, as well as eq. (\ref{loss8a}). The best-fit parameters for the
rotation-age relationship using this equation are depicted in Fig. \ref{fig2}.

Fig. \ref{fig1} clearly shows that the saturated model does not fit the data for
either type of star aged $>$ 1 Gyr. This suggests that the stars in
our sample are still under the effect of magnetic torque. Additionally,
there is a positive gradient for index $q$ from the hottest to the coolest
stars, namely from F- to G-type stars, suggesting that at a given age the coolest stars have a stronger
rotational decline, reaching the slow rotation regime faster than the hottest stars. Moreover, 
there  is also a positive gradient for braking strength $\lambda_{q}$ from the hottest to the coolest
stars. Indeed, from Table \ref{tab1} demonstrates that the present model clearly depend on stellar mass.

According to $q$ values presented in Fig. \ref{fig2},
magnetic field geometry is essentially (quasi)dipolar, irrespective of the
value of $a$. For this type of topology, our results indicate that for
F early-type stars the Pace and Pasquini law ($q=1.4$) is the most appropriate, while for late-type stars the Soderblom {\it et al.}'s
($q=2.3$) law can be considered a good approach. In case of G stars the
Skumanich ($q=3$) law is the most appropriate for early-type and the Soderblom {\it et al.} law is for late-type.

\section{Conclusions}\label{conclusions}
In summary, our study presents a new statistical approach for stellar
rotational braking. The present analysis shows that the rotation-age
relationship can
be well reproduced using a nonextensive approach from statistical
mechanics, namely the Tsallis nonextensive models. According to different studies
(\textit{e.g.:} \citet{rutten1988,matt2012}), a theoretical model for magnetic braking should
consider several parameters, ({\it e.g.} magnetic field geometry, mass loss
rate and coronal temperature), which to data are not represented in the different
laws
describing the rotation--age relations. The nonextensive analysis
proposed here considers some of these parameters, and offers the
possibility of
studying the
stellar rotational braking behaviur of different classes of stars, based on the same approach.

Our theoretical-observational results reveal that each spectral type range is characterized by a specific entropy production, this suggests the occurrence of different rotational velocity probability distribution with high energy tails as proposed by \citet{defreitas2012}. Moreover, these results suggest that increasing index $q$ behavoir between the saturated and unsaturated regimes as a function of spectral type may be due to a phase transition magnetic process that occurs in the stellar dynamo dynamics. According to \citet{bouvier1997}, low mass stars retain the memory of their initial angular momentum up to an age of about 1 Gyr, as opposed to solar type stars, where the memory is retained up to few 10$^{8}$ years. This memory can be quantified using the entropic index $q$. The fact that this index increases from F to G stars, indicates that long-range memory grows at the same scale. Higher mass stars do not maintain this type of memory for long time, because the interaction of the young star with its circumstellar disk is much shorter than their very low-mass counterparts, which retain constant angular velocity.

Finally, a comparison with observational data, namely projected
rotational velocity $v\sin i$, shows that our nonextensive approach closely fits
the decrease in stellar rotation with age for F- and G-type main--sequence
stars. Indeed, the behavior of the index $q$, which is related to the degree of 
nonextensivity, and in the present aproach can
be described by $a$ and $N$, associated to the 
dynamo process and magnetic field geometry, respectively
offers relevant information concerning the level of magnetic 
braking. At a given age the coolest stars have
a stronger rotational decline, reaching the slow rotation regime faster 
than the hottest stars. 

\section*{Acknowledgments}

Research activity of the Stellar Board of the Federal University of Rio
Grande do Norte (UFRN) is supported by continuous grants from CNPq and 
FAPERN brazilian agency. We thanks warmly an anonymous Referee for comments and suggestions that largely improved this work. We also acknowledge
continuous financial support from INCT INEspaço/CNPq/MCT.

\end{document}